\documentclass[conference,a4paper]{IEEEtran}
\usepackage{amsmath}
\usepackage{graphicx}
\usepackage{epstopdf}
\usepackage{psfrag}
\usepackage{epsfig}
\usepackage{setspace}
\usepackage{cite}

\newcommand{\ten}[1]{\overline{\overline{#1}}}

\begin{document}

\title{\huge \setstretch{0.9}Tomorrow's Metamaterials: Manipulation of \\ Electromagnetic Waves in Space, Time and Spacetime\vspace{-3mm}}

\author{\IEEEauthorblockN{Christophe Caloz, Polytechnique Montr\'{e}al, 2500 ch. de Polytechnique, H3T 1J4 Montr\'{e}al, QC, Canada}}


\maketitle

\begin{abstract}
Metamaterials represent one of the most vibrant fields of modern science and technology. They are generally dispersive structures in the direct and reciprocal space and time domains. Upon this consideration, I overview here a number of metamaterial innovations developed by colleagues and myself in the holistic framework of space and time dispersion engineering. Moreover, I provide some thoughts regarding the future perspectives of the area.
\end{abstract}

\IEEEpeerreviewmaketitle

\section{Introduction}\vspace{-1mm}

Metamaterials are electromagnetic\footnote{This represents the vast majority of metamaterials, but there are also acoustic, thermal and quantum metamaterials.} structures composed of artificial particles or ``meta-atoms'' and exhibiting properties that are typically unavailable in natural materials. In contrast to natural materials, they do no owe their properties to the actual atoms that constitute them at the angstromic scale but to their structural meta-atoms, namely to the shape, size, orientation and lattice arrangement of these meta-atoms. These properties are extremely diverse and virtually unlimited, extending well beyond the phenomenon of negative refraction that initially popularized the field. Some of these properties, and their applications, are discussed in this paper.

Historically, metamaterials are the descendants of artificial dielectrics developed by J.~C.~Bose, K.~F.~Lindman, W.~E.~Kock, W.~Rotman and others from the last decade of the nineteen's century to the 1960'ies. They really boomed at the turn of the twenty-first century, where they took their current name, with the introduction of an artificial magnetic medium based on split-ring resonators by J.~B.~Pendry and the subsequent combination of this medium with a wire electric plasma medium into a negative refractive index medium by D.~R.~Smith and colleagues. Today, metamaterials represent one of the most vibrant fields of modern science and technology. Over 100~books, among which~\cite{Caloz_Wiley_2005,Eleftheriades_Balmain_Wiley_2005,Engheta_Ziolkowski_Wiley_2006,Sarychev_Shalaev_2007,Zouhdi_Sihvola_Springer_2008,Capolino_CRC_2009,Cui_Smith_Liu_Springer_2009,Cai_Shalaev_Springer_2009,Govyadinov_VDM_2010,Maradudin_CUP_2011,Craster_Guenneau_Springer_2012,Marques_Martin_Wiley_2013,Lheurette_Wiley_2013,Werner_Kwon_Springer_2013,Solymar_Shamonina_2014}, have already been published on the topic over the past decade, and the field keeps growing steadily.

\begin{figure}[!ht]
\centering
\includegraphics[width=0.8\columnwidth]{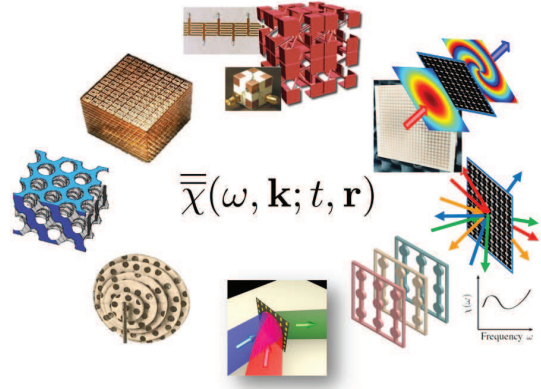}
\caption{\setstretch{1} Artistic representation of metamaterial engineering, that consists in realizing a diversity of direct and reciprocal space and time domain bianisotropic responses. Here, $\ten{\chi}$ represents the metamaterial permittivity~$(\ten{\epsilon})$, permeability~$(\ten{\mu})$, magnetic-to-electric coupling~$(\ten{\xi})$ and electric-to-magnetic coupling~$(\ten{\zeta})$ tensors, while $\omega$ is the angular frequency (reciprocal time), $\mathbf{k}$ is the spatial frequency (reciprocal space), $t$ is time (direct time) and $\mathbf{r}$ is space (direct space). Combining different dependencies $(\omega,\mathbf{k};t,\mathbf{r})$ and bi-anisotropies $(\ten{\epsilon},\ten{\mu},\ten{\xi},\ten{\zeta})$ leads to a virtually unlimited number of distinct types of metamaterials.}
\label{fig:Disp_eng_MTM}
\end{figure}

Being generally operated in temporal and spatial frequency ranges where their lattice features are sub-wavelength, metamaterials appear as homogenous media to electromagnetic waves and can therefore be characterized by effective constitutive parameters. In general, these parameters are bianisotropic, including the permittivity~($\ten{\epsilon}$), the permeability~$(\ten{\mu})$, the magnetic-to-electric coupling term~$(\ten{\xi})$ and the electric-to-magnetic coupling term~$(\ten{\zeta})$ tensors~\cite{Kong_EMW_2008}, and these tensors depend on the angular frequency~($\omega$), the spatial frequency~($\mathbf{k}$), time~($t$) and space~($\mathbf{r}$). Consequently, as illustrated in Fig.~\ref{fig:Disp_eng_MTM}, where the aforementioned tensors are represented by the generic notation $\ten{\chi}$, metamaterial engineering essentially consists in building artificial structures that realize specific direct and reciprocal space and time domain bianisotropic responses. The paper follows this logic, describing a few of the virtually unlimited metamaterial possibilities, with classification based on the $\omega,\mathbf{k},t,\mathbf{r}$ dependencies of $\ten{\epsilon},\ten{\mu},\ten{\xi},\ten{\zeta}$.

\section{Origin Spacetime Engineering}\vspace{-1mm}

\subsection{Temporal Dispersion}\vspace{-1mm}

By virtue of the second law of thermodynamics, the electromagnetic energy stored in a system is necessarily a positive quantity. As a consequence, all media, except vacuum, are frequency dispersive, i.e. have $\ten{\chi}=\ten{\chi}(\omega)$~\cite{Landau_BH_1984,Schwinger_WP_1998}. Metamaterials tend to be particularly strongly dispersive, with a level of dispersion that is proportional to how much they alter waves, according to causality principles, incarnated by Kramers-Kronig relations~\cite{Jackson_Wiley_1998,Landau_BH_1984,Schwinger_WP_1998}.

\subsection{Coupled Resonator Broad Bandwidth}\vspace{-1mm}

Two types of metamaterial structures have been often opposed in the early days of ``modern metamaterials'', i.e. in the first decade of the century: wire/split-ring and transmission-line metamaterials~\cite{Caloz_Wiley_2005,Eleftheriades_Balmain_Wiley_2005,Engheta_Ziolkowski_Wiley_2006}. The fundamental nature of the difference was realized only later.

The wire/split-ring metamaterials, mostly developed by physicists, were conceived as natural matter, i.e. as arrays of isolated atoms, whose Lorentz dispersive response is obtained by applying Newton equation of motion to an isolated atom and then averaging equivalent electric and magnetic dipole moments into corresponding polarization densities~\cite{Jackson_Wiley_1998}. These metamaterials are thus essentially systems of \emph{uncoupled resonators} and their bandwidth is therefore limited by their quality factor, where low loss is inevitably accompanied with narrow bandwidth.

In contrast, transmission-line metamaterials, mostly developed by engineers, are \emph{coupled resonator} structures. The resonators that constitute them are essentially similar to those of the wire/split-ring metamaterials, but they are tightly packed together and hence exchange energy. This makes all the difference: the coupled states\footnote{An isolated resonator is characterized by its resonant mode, or state, in a given frequency range. When two identical and initially distant resonators are brought in close proximity to each other, a new structure is formed, with richer field distributions (typically symmetric and anti-symmetric) and the initially degenerate states of the isolated resonators split into two coupled states. Three coupled resonators leads to three coupled states, and so on.} of the numerous meta-atom resonators collapse into a continuum, and this gives rise to extremely broad bandwidths without penalty on transmission\footnote{The concept of coupled resonators for broad bandwidth is well known in microwave filter theory\cite{Matthaei_MGH_1964}.}~\cite{Caloz_MT_03_2009}.

Composite right/left-handed (CRLH) transmission line metamaterials and derivatives~\cite{Rennings_APMC_12_2006,Caloz_MWCL2_11_2006,Caloz_PSSb_4_2007}, that may be seen as a coupled-resonator extension of Lorentz-Drude media with double negative and positive parameters below and above their balanced electric and magnetic plasma or zero index frequency, have lead to a myriad of novel component and antenna applications~\cite{Caloz_APS_06_2002,Caloz_MWCL_12_2003,Lai_MM_09_2004,Caloz_MOTL_03_2004,Caloz_Wiley_2005,Caloz_NJP_08_2005,Nguyen_TMTT_5_2007,Zedler_TMTT_12_2007,Caloz_APM_10_2008,Caloz_MT_03_2009}. Ironically, whereas some physicists deployed enormous efforts to overcome the loss-bandwidth issue of wire/split-ring metamaterials, engineers routinely used CRLH structures to enhance or multiplicate the bandwidth of existing components!

\subsection{From Dispersion and Bandwidth to Spacetime Engineering}\vspace{-1mm}

It soon occurred to me that the combination of high dispersion, inherent to metamaterials, and broad bandwidth, available in coupled-resonator metamaterials, represented a unique opportunity for \emph{dispersion engineering}, i.e. for developing novel media with tailorable frequency dependent parameters, $\ten{\chi}=\ten{\chi}(\omega)$, and that this would lead to a new range of applications. In this perspective, I showed in~\cite{Caloz_PIEEE_10_2011} how one could leverage the high dispersion and broad bandwidth of metamaterials to engineer artificial phase velocity, group velocity and group velocity dispersion parameters for corresponding applications, and that these possibilities could be further extended by using multi-scale structures such as ferromagnetic nanowire metamaterials~\cite{Carignan_APL_08_2009,Carignan_TMTT_10_2011}.

The aforementioned dispersion engineering concepts are purely temporal. However, metamaterials, as their photonic crystal predecessors, are also generally spatially dispersive, i.e. $\ten{\chi}=\ten{\chi}(\mathbf{k})$. For instance, a chiral metamaterial, that reciprocally rotates the polarization of waves, is inherently spatially dispersive\footnote{Consider the general electric response of a bi-isotropic medium to a plane wave: \mbox{$\mathbf{D}=\epsilon\mathbf{E}+\xi\mathbf{H}
=\epsilon\mathbf{E}+j\xi/(\omega\mu)\nabla\times\mathbf{E}
=\epsilon\mathbf{E}+\xi/(\omega\mu)\mathbf{k}\times\mathbf{E}$}. So, the medium may be considered as a dielectric medium with effective permittivity tensorial operator $\ten{\epsilon}_\text{e}=\epsilon\ten{I}+\xi/(\omega\mu)\mathbf{k}\times
=\ten{\epsilon}_\text{e}(\mathbf{k})$.}. A generalized-refractive metasurface is also a spatially dispersive structure since it transforms incident $\mathbf{k}$ vectors to different reflected and transmitted $\mathbf{k}$ vectors. So, metamaterials are in general both temporally and spatially dispersive, $\ten{\chi}=\ten{\chi}(\omega,\mathbf{k})$.

A large number of metamaterials are also nonuniform, i.e. having properties changing in space. This is for instance the case of artificial Luneburg or fisheye lenses and coordinate-transform cloaking metamaterials. So, the parameters of metamaterials generally also depend on space, $\ten{\chi}=\ten{\chi}(\omega,\mathbf{k};\mathbf{r})$.

Time is apparently missing in the zoo of metamaterials formed so far. Adding time, i.e. $\ten{\chi}=\ten{\chi}(\omega,\mathbf{k};t,\mathbf{r})$, corresponds to making the metamaterial either moving or, more practically, pumped. Such metamaterials have been hardly studied so far, and I shall also say a few words about them near the end of the paper.

We have thus established a general direct and reciprocal space and time -- or generally \emph{spacetime} -- classification and description of metamaterials, as illustrated in Fig.~\ref{fig:Disp_eng_MTM}.

\section{Temporal Frequency Engineering -- $\ten{\chi}(\omega)$}\vspace{-1mm}

Current technology cannot meet the ever increasing demand in faster, more reliable and ubiquitous radio systems. The trend over past decades has been to place an ever increasing emphasis on digital signal processing (DSP), where spectral efficiency has been maximized via sophisticated modulation, multiplexing or MIMO schemes. However, these approaches have now reached their limits and migration to millimeter-wave and terahertz frequencies has become indispensable to access larger spectral resources. Unfortunately, DSP is not applicable at such frequencies, where signals are varying too fast to be digitized. To overcome this problem, my group, inspired by the capabilities of broadband metamaterials~\cite{Caloz_PIEEE_10_2011} and by concepts of ultrafast optics~\cite{Saleh_Teich_2007}, recently proposed real-time analog signal processing (R-ASP) as a new and general technology for processing electromagnetic waves in real time~\cite{Caloz_MM_RASP_Caloz_09_2013}.

The core of a R-ASP system is the~\emph{phaser}, which is a dispersive component following specified group delay versus frequency responses for diverse operations, such as for instance real-time Fourier transformation. The initial approach for the design of phasers was based on CRLH meta-lines but was later extended to more general dispersive structures, that attained unprecedented level of temporal frequency control at radio frequencies. A diversity of phasers, in various technologies -- transmission and reflection, planar, multilayer and waveguide, passive and active, reciprocal and nonreciprocal, guided or spatial technologies -- were reported~\cite{Coulombe_TMTT_08_2009,Gupta_TMTT_09_2010,Nikfal_TMTT_06_2011,Gupta_TMTT_12_2012,Zhang_APM_04_2013,Zhang_EL_07_2013,Zou_arXiv_04_2014,Gupta_TMTT_03_2015,Liao_MWCL_06_2015,Zhang_TMTT_09_2015,Guo_MWCL_12_2015} with efficient synthesis techniques~\cite{Horii_MWCL_01_2012,Zhang_TMTT_08_2012,Zhang_TMTT_03_2013,Gupta_IJCTA_05_2013,Zhang_TMTT_12_2013,Zhang_IJRMCAE_05_2014}. Related applications included tunable pulse delay lines, pulse position modulators, compressive transceivers, frequency discriminators, temporal RFIDs, dispersion code multiple access (DCMA) systems, spread spectrum transmitters and time-reversal subwavelength transmission~\cite{Abielmona_MWCL_12_2007,Nguyen_MWCL_08_2008,Abielmona_TTMTT_11_2009,Nikfal_TMTT_06_2011,Gupta_AWPL_11_2011,Nikfal_MWCL_11_2012,Nikfal_11_2013,Nikfal_MWCL_11_2014,Ding_TAP_09_2015}.

A major challenge in real-time processing is the fact that, due to fundamental causality constraints, manipulating the frequency response of a medium or device generally implies variations of the magnitude of the transfer function. However, we recently reported a fundamental concept that overcomes this limitation, and hence opens up new horizons in R-ASP. This is the concept of a \emph{perfect dispersive medium}. This is a metamaterial, composed of loss-gain unit cells with properly tuned electric and magnetic dipolar responses, that exhibits a perfectly flat magnitude response with arbitrary (within causality) dispersion (phase) response~\cite{Gupta_AXV1511_11_2015}. This medium also represents a \emph{spatial phaser}, transforming the temporal spectrum of electromagnetic waves in quasi-arbitrary fashions.

\section{Spatial Frequency Engineering -- $\ten{\chi}(\mathbf{k})$}\vspace{-1mm}

The first metamaterials, including artificial dielectric lenses and negative refraction media, were designed to operate in the monochromatic regime, i.e. $\ten{\chi}\neq\ten{\chi}(\omega)$, transforming the spatial spectrum of electromagnetic waves and hence exhibiting spatial frequency variations, i.e. $\ten{\chi}=\ten{\chi}(\mathbf{k})$, as typical Fourier transforming devices~\cite{Goodman_2004}. However, the most powerful spatial-frequency engineered metamaterials nowadays are metasurfaces.

Metasurface may be seen as the two-dimensional counterparts of volume metamaterials and the functional extensions of frequency and polarization selective surfaces. Arguably, they possess the benefits of both without suffering from their drawbacks, hence leading to an incredible wealth of electromagnetic transformations, reported by many groups around the world and including holographic beam forming, generalized refraction, fast imaging, vortex generation, angular filtering and mathematical processing. Note that metasurfaces are also most often spatially varying, i.e. $\ten{\chi}=\ten{\chi}(\mathbf{k};\mathbf{r})$, where the spatial and direct space descriptions are simply related by Fourier pair transformations.

A daunting question just a few years ago was: how can one synthesize a general metasurface that would transform an incident wave into specified scattered (reflected and transmitted) waves? We initially developed a momentum conservation synthesis method in~\cite{Salem_OE_06_2014}, based on the fact that momentum ($\mathbf{k}$) must be conserved across a metasurface due to its zero (or deeply subwavelength) thickness, but this method was overly complicated for vectorial waves, where it involved field expansions in vectorial Bessel functions. We subsequently developed a surface susceptibility synthesis method, based on generalized sheet transition conditions (GSTCs), that not only powerfully transforms any incident electromagnetic waves into arbitrary reflected and transmitted waves, but also provides deep insight in the physics of the transformation, and revealed the possibility to simultaneously perform multiple transformations leveraging bianisotropic susceptibility degrees of freedom~\cite{Achouri_TMTT_07_2015,Salem_PIER_11_2014}. We subsequently reported the following applications: a generalized refraction beam splitter based on refractive birefringence, and double orbital angular momentum generator, a spatial switch, and a spatial processor that may be seen as the electromagnetic extension of both an interferometer and a transistor~\cite{Achouri_AXV1510_10_2015_2,Achouri_AXV1510_1_10_2015}. 

In connection with the aforementioned research, we realized that nonreciprocity, breaking the symmetry of susceptibility tensors and hence liberating extra degrees of freedom, was a tool for \emph{transformation diversity}, in addition of its applications to isolating and circulating devices. However, all nonreciprocal materials reported until recently were based on ferromagnetic substances and hence require a biasing magnet that would electromagnetically obstruct the aperture of the metasurface, hence making it unusable. Fortunately, we had just discovered, in 2011, what I like to call ``magnetless magnetism'', namely the possibility to produce all the magnetic phenomena with a magnetless metamaterial structure, whose transistor loaded particles mimic electron spin precession in natural magnetic materials~\cite{Kodera_APL_07_2011,Sounas_TAP_01_2013}. This metamaterial has already been demonstrated in several component applications that would normally require magnetic materials~\cite{Kodera_AWPL_01_2012,Kodera_AWPL_12_2012,Kodera_TMTT_03_2013}. In addition, they lend themselves well to metasurface configurations, and may therefore indeed provide the aforementioned transformation diversity in future sophisticated metasurfaces.

\section{Spacetime Frequency Engineering -- $\ten{\chi}(\omega;\mathbf{k})$}\vspace{-1mm}

Many natural phenomena (e.g. rainbows or colorful oil slicks) and classical human devices (e.g. Newton prisms, diffraction gratings, or color holograms) are based on spectral decomposition, namely the mapping of temporal frequencies ($\omega$) into spatial frequencies ($\mathbf{k}$). In fact, a leaky-wave antenna, excited by a pulse rather than a monochromatic wave, effectively operates as a diffraction grating. Based on this observation, a unique real-time leaky-wave antenna based spectrogram analyzer, scalable to arbitrary frequencies, was reported in~\cite{Gupta_TMTT_04_2009}. This application was enabled by the full-space scanning capability introduced by the CRLH leaky-wave antenna~\cite{Liu_EL_11_2002,Caloz_APM_10_2008}, that, incidentally, allowed to solve the persistent problem of broadside radiation instability previously plaguing also other leaky-wave antennas~\cite{Otto_TAP_10_2011,Otto_AWPL_06_2012,Otto_TAP_04_2014,Otto_TAP_10_2014}, and led to several related novel applications and technologies~\cite{Casares_TAP_08_2006,Caloz_APM_10_2008,Kodera_TMTT_04_2009,Nguyen_TMTT_07_2010,Kodera_TAP_08_2010,Kodera_TAP_10_2010,Abielmona_TAP_04_2011,Kodera_AWPL_01_2012}. However, spatiotemporal, or spacetime, engineering may be pushed further. For instance, leaky-wave antenna spectral decomposition may be used for the real-time spectral analysis of ultra-fast signals where different frequency components are mapped onto a screen with related photo-detectors~\cite{Gutpa_APS_07_2015_2}. The resolution of such a device, for a given projection depth, could be dramatically increased if the mapping could be performed on a two-dimensional screen of detectors, as suggested in~\cite{Gupta_arXiv7791_12_2014}, using a sophisticated metasurface, involving spatial phasing, gradient profile and multiple resonances.

\section{Spacetime Engineering -- $\ten{\chi}(t,\mathbf{r})$}\vspace{-1mm}

Direct space varying metamaterials, $\ten{\chi}=\ten{\chi}(\mathbf{r})$, are gradient metamaterials, already touched upon in previous sections. The time dependence is much less obvious and has been hardly approached to date in the realm of metamaterials. A time-dependent metamaterial is a time-varying structure, $\ten{\chi}=\ten{\chi}(t)$ -- sort of a ``living metamaterial'' -- whose properties change in time, and, most interestingly, in a diabatic\footnote{``Diabatic'' is meant here as very fast, specifically sub-period, where the period is defined as that of the highest frequency component of the spectrum.} fashion, so as to allow transformations of the temporal spectrum of waves. Such metamaterials can be of two different types: it may be a moving medium~\cite{Kong_EMW_2008} or a parametric~\cite{Collin_1992} structure. The former is generally unpractical, but the later is possess a huge innovation potential, in my opinion.

There exists a partial duality between spatially and temporally varying, or discontinuous, media~\cite{Salem_AXV02012_05_2015}. Both obey, at least in one dimension, similar wave equations and boundary conditions, but the latter does not allow back-scattering because reflection back to the past are causally impossible. This partial duality leads to fascinating novel physical and technological concepts. Let us consider just one example for the sake of illustration: the spacetime modulated leaky-wave system reported last year in~\cite{Taravati_APS_07_2015}. Operating oblique transitions in the dispersion diagram, as silicon in its indirect bandgap, the system is at the same time an up-converter, a transmit antenna, a nonreciprocal duplexer, a receiver and a down-converter!

\section{Future}\vspace{-1mm}

The twenty-first century, considering recent spectacular progress in nanotechnologies, will doubtlessly witness a metamaterial revolution. Given the explosion of biotechnologies in these times~\cite{Dyson_NRB_2015}, it is highly probable that some of the most advanced metamaterials will involve biological material, as DNA. Conferences in a couple of decades may feature special sessions on genetically engineered spacetime metamaterials.

\bibliography{2016_04_EuCAP_DAVOS_ST_MTMS_INVITED_Caloz}

\begin{thebibliography}{10}
\providecommand{\url}[1]{#1}
\csname url@samestyle\endcsname
\providecommand{\newblock}{\relax}
\providecommand{\bibinfo}[2]{#2}
\providecommand{\BIBentrySTDinterwordspacing}{\spaceskip=0pt\relax}
\providecommand{\BIBentryALTinterwordstretchfactor}{4}
\providecommand{\BIBentryALTinterwordspacing}{\spaceskip=\fontdimen2\font plus
\BIBentryALTinterwordstretchfactor\fontdimen3\font minus
  \fontdimen4\font\relax}
\providecommand{\BIBforeignlanguage}[2]{{%
\expandafter\ifx\csname l@#1\endcsname\relax
\typeout{** WARNING: IEEEtran.bst: No hyphenation pattern has been}%
\typeout{** loaded for the language `#1'. Using the pattern for}%
\typeout{** the default language instead.}%
\else
\language=\csname l@#1\endcsname
\fi
#2}}
\providecommand{\BIBdecl}{\relax}
\BIBdecl

\bibitem{Caloz_Wiley_2005}
C.~Caloz and T.~Itoh, \emph{Electromagnetic Metamaterials, Transmission Line
  Theory and Microwave Applications}.\hskip 1em plus 0.5em minus 0.4em\relax
  \bstyle{Wiley - IEEE Press}, 2005.

\bibitem{Eleftheriades_Balmain_Wiley_2005}
G.~V. Eleftheriades and K.~G. Balmain, Eds., \emph{Negative-Refraction
  Metamaterials: Fundamental Principles and Applications}.\hskip 1em plus 0.5em
  minus 0.4em\relax \bstyle{Wiley - IEEE Press}, 2005.

\bibitem{Engheta_Ziolkowski_Wiley_2006}
N.~Engheta and R.~Ziolkowski, Eds., \emph{Metamaterials: Physics and
  Engineering Explorations}.\hskip 1em plus 0.5em minus 0.4em\relax
  \bstyle{Wiley - IEEE Press}, 2006.

\bibitem{Sarychev_Shalaev_2007}
A.~K. Saruchev and V.~M. Shalaev, \emph{Electrodynamics Of
  Metamaterials}.\hskip 1em plus 0.5em minus 0.4em\relax Wspc, 2007.

\bibitem{Zouhdi_Sihvola_Springer_2008}
S.~Zouhdi and A.~Sihvola, Eds., \emph{Metamaterials and Plasmonics:
  Fundamentals, Modelling, Applications}.\hskip 1em plus 0.5em minus
  0.4em\relax Springer, 2008.

\bibitem{Capolino_CRC_2009}
F.~Capolino, Ed., \emph{Metamaterials Handbook}.\hskip 1em plus 0.5em minus
  0.4em\relax CRC Press, 2009, 2 volumes.

\bibitem{Cui_Smith_Liu_Springer_2009}
T.~J. Cui, D.~Smith, and R.~Liu, Eds., \emph{Metamaterials: Theory, Design, and
  Applications}.\hskip 1em plus 0.5em minus 0.4em\relax Springer, 2009.

\bibitem{Cai_Shalaev_Springer_2009}
W.~Cai and V.~Shalaev, \emph{Optical Metamaterials: Fundamentals and
  Applications}.\hskip 1em plus 0.5em minus 0.4em\relax Springer, 2009.

\bibitem{Govyadinov_VDM_2010}
A.~Govyadinov, \emph{Light Propagation in Nanostructured Metamaterials and
  Micro-Resonators}.\hskip 1em plus 0.5em minus 0.4em\relax VDM Verlag Dr.
  M\"{u}ller, 2010.

\bibitem{Maradudin_CUP_2011}
A.~A. Marududin, Ed., \emph{Structured Surfaces as Optical
  Metamaterials}.\hskip 1em plus 0.5em minus 0.4em\relax Cambridge University
  Press, 2011.

\bibitem{Craster_Guenneau_Springer_2012}
R.~V. Craster and S.~Guenneau, Eds., \emph{Acoustic Metamaterials: Negative
  Refraction, Imaging, Lensing and Cloaking}.\hskip 1em plus 0.5em minus
  0.4em\relax Springer, 2012.

\bibitem{Marques_Martin_Wiley_2013}
R.~Marqu\'{e}s and F.~Mart\'{\i}n, \emph{Metamaterials with Negative
  Parameters: Theory, Design and Microwave Applications}.\hskip 1em plus 0.5em
  minus 0.4em\relax Wiley-Interscience, 2013.

\bibitem{Lheurette_Wiley_2013}
E.~Lheurette, \emph{Metamaterials and Wave Control}.\hskip 1em plus 0.5em minus
  0.4em\relax Wiley-ISTE, 2013.

\bibitem{Werner_Kwon_Springer_2013}
D.~H. Werner and D.-H. Kwon, \emph{Transformation Electromagnetics and
  Metamaterials: Fundamental Principles and Applications}.\hskip 1em plus 0.5em
  minus 0.4em\relax Springer, 2013.

\bibitem{Solymar_Shamonina_2014}
L.~Solymar and E.~Shamonina, \emph{Waves in Metamaterials}.\hskip 1em plus
  0.5em minus 0.4em\relax Oxford University Press, 2014.

\bibitem{Kong_EMW_2008}
J.~A. Kong, \emph{Electromagnetic Wave Theory}.\hskip 1em plus 0.5em minus
  0.4em\relax EMW Publishing, 2008.

\bibitem{Landau_BH_1984}
L.~D. Landau and L.~P. Pitaevskii, \emph{Electrodynamics of Continuous Media},
  2nd~ed.\hskip 1em plus 0.5em minus 0.4em\relax Butterworth-Heinemann, 1984.

\bibitem{Schwinger_WP_1998}
J.~Schwinger, L.~L. Deraad, K.~Milton, W.-Y. Tsai, and J.~Norton,
  \emph{Classical Electrodynamics}.\hskip 1em plus 0.5em minus 0.4em\relax
  Westview Press, 1998.

\bibitem{Jackson_Wiley_1998}
J.~D. Jackson, \emph{Classical Electrodynamics}, 3rd~ed.\hskip 1em plus 0.5em
  minus 0.4em\relax Wiley, 1998.

\bibitem{Matthaei_MGH_1964}
G.~L. Matthaei, L.~Young, and E.~M.~T. Jones, \emph{Microwave filters,
  impedance-matching networks, and coupling structures}.\hskip 1em plus 0.5em
  minus 0.4em\relax McGraw-Hill, 1964.

\bibitem{Caloz_MT_03_2009}
C.~Caloz, ``Perspectives on {EM} metamaterials,'' \emph{\jstyle{Materials
  Today}}, vol.~12, no.~3, pp. 12--20, Mar. 2009, \istyle{invited}.

\bibitem{Rennings_APMC_12_2006}
A.~Rennings, S.~Otto, J.~Mosig, C.~Caloz, and I.~Wolff, ``Extended composite
  right/left-handed ({E-CRLH}) metamaterial and its application as quadband
  quarter-wavelength transmission line,'' in \emph{\cstyle{IEEE Asia Pacific
  Microw. Conf. (APMC)}}, Yokohama, Japan, Dec. 2006.

\bibitem{Caloz_MWCL2_11_2006}
C.~Caloz, ``Dual composite right/left-handed ({D-CRLH}) transmission line
  metamaterial,'' \emph{\jstyle{IEEE Microw. Wireless Compon. Lett.}}, vol.~16,
  no.~11, pp. 585--587, Nov. 2006.

\bibitem{Caloz_PSSb_4_2007}
C.~Caloz, S.~Abielmona, H.~V. Nguyen, and A.~Rennings, ``Dual composite
  right/left-handed ({D-CRLH}) leaky-wave antenna with low beam squinting and
  tunable group velocity,'' \emph{\jstyle{Phys. Stat. Solidi (b)}}, vol. 244,
  no.~4, pp. 1219--1226, Apr. 2007, \istyle{invited}.

\bibitem{Caloz_APS_06_2002}
C.~Caloz and T.~Itoh, ``Application of the transmission line theory of
  left-handed ({LH}) materials to the realization of a microstrip lh
  transmission line,'' in \emph{\cstyle{IEEE AP-S Int. Antennas Propag.
  (APS)}}, San Antonio, TX, Jun. 2002, pp. 412--415, \istyle{invited}.

\bibitem{Caloz_MWCL_12_2003}
------, ``Positive / negative refractive index anisotropic {2D}
  metamaterials,'' \emph{\jstyle{IEEE Microw. Wireless Compon. Lett.}},
  vol.~13, no.~12, pp. 547--549, Dec. 2003.

\bibitem{Lai_MM_09_2004}
A.~Lai, C.~Caloz, and T.~Itoh, ``Composite right/left-handed transmission line
  metamaterials,'' \emph{\jstyle{IEEE Microw. Mag.}}, vol.~5, no.~3, pp.
  34--50, Sept. 2004, \istyle{invited}.

\bibitem{Caloz_MOTL_03_2004}
C.~Caloz, I.-H. Lin, and T.~Itoh, ``Characteristics and potential applications
  of nonlinear left-handed transmission lines,'' \emph{\jstyle{Microw. Opt.
  Technology Lett.}}, vol.~40, no.~6, pp. 471--473, Mar. 2004.

\bibitem{Caloz_NJP_08_2005}
C.~Caloz, A.~Lai, and T.~Itoh, ``The challenge of homogenization in
  metamaterials,'' \emph{\jstyle{New J. Phys.}}, vol.~7, no. 167, pp. 1--15,
  Aug. 2005.

\bibitem{Nguyen_TMTT_5_2007}
H.~V. Nguyen and C.~Caloz, ``Generalized coupled-mode approach of metamaterial
  coupled-line couplers: complete theory, explanation of phenomena and
  experimental demonstration,'' \emph{\jstyle{IEEE Trans. Microw. Theory
  Tech.}}, vol.~55, no.~5, pp. 1029--1039, May 2007.

\bibitem{Zedler_TMTT_12_2007}
M.~Zedler, C.~Caloz, and P.~Russer, ``A 3-{D} isotropic left-handed
  metamaterial based on the rotated transmission-line matrix ({TLM}) scheme,''
  \emph{\jstyle{IEEE Trans. Microw. Theory Tech.}}, vol.~55, no.~12, pp.
  2930--2941, Dec. 2007.

\bibitem{Caloz_APM_10_2008}
C.~Caloz, T.~Itoh, and A.~Rennings, ``{CRLH} metamaterial leaky-wave and
  resonant antennas,'' \emph{\jstyle{IEEE Antennas Propag. Mag.}}, vol.~50,
  no.~5, pp. 25--39, Oct. 2008.

\bibitem{Caloz_PIEEE_10_2011}
C.~Caloz, ``Metamaterial dispersion engineering concepts and applications,''
  \emph{\jstyle{Proc. IEEE}}, vol.~99, no.~10, pp. 1711--1719, Oct. 2011.

\bibitem{Carignan_APL_08_2009}
L.-P. Carignan, V.~Boucher, T.~Kodera, C.~Caloz, A.~Yelon, and D.~M\'{e}nard,
  ``Double ferromagnetic resonance in nanowire arrays,'' \emph{\jstyle{Appl.
  Phys. Lett.}}, vol.~45, no.~6, pp. 062\,504--1:3, Aug. 2009.

\bibitem{Carignan_TMTT_10_2011}
L.-P. Carignan, A.~Yelon, D.~M\'{e}nard, and C.~Caloz, ``Ferromagnetic nanowire
  metamaterials: theory and applications,'' \emph{\jstyle{IEEE Trans. Microw.
  Theory Tech.}}, vol.~59, no.~19, pp. 2568--2586, Oct. 2011.

\bibitem{Saleh_Teich_2007}
B.~E.~A. Saleh and M.~C. Teich, \emph{Fundamentals of Photonics}, 2nd~ed.\hskip
  1em plus 0.5em minus 0.4em\relax Wiley, 1964, chap.~22.

\bibitem{Caloz_MM_RASP_Caloz_09_2013}
C.~Caloz, S.~Gupta, Q.~Zhang, and B.~Nikfal, ``Analog signal processing,''
  \emph{\jstyle{IEEE Microw. Mag.}}, vol.~14, no.~6, pp. 87--103, Sept. 2013,
  \istyle{invited}.

\bibitem{Coulombe_TMTT_08_2009}
M.~Coulombe and C.~Caloz, ``Reflection-type artificial dielectric substrate
  microstrip dispersive delay line ({DDL}) for analog signal processing,''
  \emph{\jstyle{IEEE Trans. Microw. Theory Tech.}}, vol.~57, no.~7, pp.
  1714--1723, Jul. 2009.

\bibitem{Gupta_TMTT_09_2010}
S.~Gupta, A.~Parsa, E.~Perret, R.~V. Snyder, R.~J. Wenzel, and C.~Caloz,
  ``Group delay engineered non-commensurate transmission line all-pass network
  for analog signal processing,'' \emph{\jstyle{IEEE Trans. Microw. Theory
  Tech.}}, vol.~58, no.~9, pp. 2392--2407, Sept. 2010.

\bibitem{Nikfal_TMTT_06_2011}
B.~Nikfal, S.~Gupta, and C.~Caloz, ``Increased group delay slope loop system
  for enhanced-resolution analog signal processing,'' \emph{\jstyle{IEEE Trans.
  Microw. Theory Tech.}}, vol.~59, no.~6, pp. 1622--1628, Jun. 2011.

\bibitem{Gupta_TMTT_12_2012}
S.~Gupta, D.~L. Sounas, H.~V. Nguyen, Q.~Zhang, and C.~Caloz, ``{CRLH-CRLH}
  {C-section} dispersive delay structures with enhanced group delay swing for
  higher analog signal processing resolution,'' \emph{\jstyle{IEEE Trans.
  Microw. Theory Tech.}}, vol.~60, no.~12, pp. 3939--3949, Dec. 2012.

\bibitem{Zhang_APM_04_2013}
Q.~Zhang, D.~L. Sounas, S.~Gupta, and C.~Caloz, ``Wave interference explanation
  of group delay dispersion in resonators,'' \emph{\jstyle{IEEE Antennas
  Propag. Mag.}}, vol.~55, no.~2, pp. 212--227, Apr. 2013.

\bibitem{Zhang_EL_07_2013}
Q.~Zhang and C.~Caloz, ``Comparison of transmission and reflection allpass
  phasers for analog signal processing,'' \emph{\jstyle{Electron. Lett.}},
  vol.~49, no.~14, Jul. 2013.

\bibitem{Zou_arXiv_04_2014}
L.~Zou, Q.~Zhang, and C.~Caloz, ``Planar reflective phaser and synthesis for
  radio analog signal processing {R-ASP},'' \emph{\jstyle{arXiv}:1404.2628},
  Apr. 2014.

\bibitem{Gupta_TMTT_03_2015}
S.~Gupta, Q.~Zhang, L.~Zou, L.~J. Kiang, and C.~Caloz, ``Generalized
  coupled-line all-pass phasers,'' \emph{\jstyle{IEEE Trans. Microw. Theory
  Tech.}}, vol.~63, no.~3, pp. 1007--1018, Mar. 2015.

\bibitem{Liao_MWCL_06_2015}
W.~Liao, Q.~Zhang, Y.~Chen, and C.~Caloz, ``Compact reflection-type phaser
  using quarter-wavelength transmission line resonators,'' \emph{\jstyle{IEEE
  Trans. Antennas Propag.}}, vol.~25, no.~6, pp. 391--393, Jun. 2015.

\bibitem{Zhang_TMTT_09_2015}
Q.~Zhang, T.~Guo, B.~A. Khan, T.~Kodera, and C.~Caloz, ``Coupling matrix
  synthesis of nonreciprocal lossless two-port networks using gyrators and
  inverters,'' \emph{\jstyle{IEEE Trans. Microw. Theory Tech.}}, vol.~63,
  no.~9, pp. 2782--2792, Sept. 2015.

\bibitem{Guo_MWCL_12_2015}
T.~Guo, Q.~Zhang, Y.~Chen, R.~Zhang, and C.~Caloz, ``Single-step tunable group
  delay phaser for real-time spectrum sniffing,'' \emph{\jstyle{IEEE Microw.
  Wireless Compon. Lett.}}, vol.~25, no.~12, pp. 808--810, Dec. 2015.

\bibitem{Horii_MWCL_01_2012}
Y.~Horii, S.~Gupta, B.~Nikfal, and C.~Caloz, ``Multilayer broadside-coupled
  dispersive delay structures for analog signal processing,''
  \emph{\jstyle{IEEE Microw. Wireless Compon. Lett.}}, vol.~22, no.~1, pp.
  1--3, Jan. 2012.

\bibitem{Zhang_TMTT_08_2012}
Q.~Zhang, S.~Gupta, and C.~Caloz, ``Synthesis of narrow-band reflection-type
  phaser with arbitrary prescribed group delay,'' \emph{\jstyle{IEEE Trans.
  Microw. Theory Tech.}}, vol.~60, no.~8, pp. 2394--2402, Aug. 2012.

\bibitem{Zhang_TMTT_03_2013}
Q.~Zhang, D.~L. Sounas, and C.~Caloz, ``Synthesis of cross-coupled
  reduced-order dispersive delay structures ({DDS}) with arbitrary group delay
  and controlled magnitude,'' \emph{\jstyle{IEEE Trans. Microw. Theory Tech.}},
  vol.~61, no.~3, pp. 1043--1052, Mar. 2013.

\bibitem{Gupta_IJCTA_05_2013}
S.~Gupta, D.~L. Sounas, Q.~Zhang, and C.~Caloz, ``All-pass dispersion synthesis
  using microwave {C-sections},'' \emph{\jstyle{Int. J. Circ. Theory Appl.}},
  pp. 1--18, May 2013.

\bibitem{Zhang_TMTT_12_2013}
Q.~Zhang, J.~W. Bandler, and C.~Caloz, ``Design of dispersive delay structures
  ({DDS}s) formed by coupled {C}-sections using predistortion with space
  mapping,'' \emph{\jstyle{IEEE Trans. Microw. Theory Tech.}}, vol.~61, no.~12,
  pp. 4040--4051, Dec. 2013.

\bibitem{Zhang_IJRMCAE_05_2014}
Q.~Zhang, S.~Gupta, and C.~Caloz, ``Synthesis of broadband phasers formed by
  commensurate {C- and D-sections},'' \emph{\jstyle{Int. J. RF Microw. Comput.
  Aided Eng.}}, vol.~24, no.~3, pp. 322--331, May 2014.

\bibitem{Abielmona_MWCL_12_2007}
S.~Abielmona, S.~Gupta, and C.~Caloz, ``Experimental demonstration and
  characterization of a tunable {CRLH} delay line system for impulse/continuous
  wave,'' \emph{\jstyle{IEEE Microw. Wireless Compon. Lett.}}, vol.~17, no.~12,
  pp. 864--866, Dec. 2007.

\bibitem{Nguyen_MWCL_08_2008}
H.~V. Nguyen and C.~Caloz, ``Composite right/left-handed delay line pulse
  position modulation transmitter,'' \emph{\jstyle{IEEE Microw. Wireless
  Compon. Lett.}}, vol.~18, no.~5, pp. 527--529, Aug. 2008.

\bibitem{Abielmona_TTMTT_11_2009}
S.~Abielmona, S.~Gupta, and C.~Caloz, ``Compressive receiver using a
  {CRLH}-based dispersive delay line for analog signal processing,''
  \emph{\jstyle{IEEE Trans. Microw. Theory Tech.}}, vol.~57, no.~11, pp.
  2617--2626, Nov. 2009.

\bibitem{Gupta_AWPL_11_2011}
S.~Gupta, B.~Nikfal, and C.~Caloz, ``Chipless {RFID} system based on group
  delay engineered dispersive delay structures,'' \emph{\jstyle{IEEE Antennas
  Wirel. Propag. Lett.}}, vol.~10, pp. 1366--1368, Dec. 2011.

\bibitem{Nikfal_MWCL_11_2012}
B.~Nikfal, D.~Badiere, M.~Repeta, B.~Deforge, S.~Gupta, and C.~Caloz,
  ``Distortion-less real-time spectrum sniffing based on a stepped group-delay
  phaser,'' \emph{\jstyle{IEEE Microw. Wireless Compon. Lett.}}, vol.~22,
  no.~11, pp. 601--603, Nov. 2012.

\bibitem{Nikfal_11_2013}
B.~Nikfal, M.~Salem, and C.~Caloz, ``A method and apparatus for encoding data
  using instantaneous frequency dispersion,'' \emph{US 62/002,978}, Nov. 2013.

\bibitem{Nikfal_MWCL_11_2014}
B.~Nikfal, Q.~Zhang, and C.~Caloz, ``Enhanced-{SNR} impulse radio transceiver
  based on phasers,'' \emph{\jstyle{IEEE Microw. Wireless Compon. Lett.}},
  vol.~24, no.~11, pp. 778--790, Nov. 2014.

\bibitem{Ding_TAP_09_2015}
S.~Ding, R.~Zang, L.~Zou, B.~Wang, and C.~Caloz, ``Enhancement of time-reversal
  subwavelength wireless transmission using pulse shaping,'' \emph{\jstyle{IEEE
  Trans. Antennas Propag.}}, vol.~63, no.~9, pp. 4169--4174, Sept. 2015.

\bibitem{Gupta_AXV1511_11_2015}
S.~Gupta and C.~Caloz, ``Perfect dispersive medium,''
  \emph{\jstyle{arXiv}:1511.00671}, Nov. 2015.

\bibitem{Goodman_2004}
J.~W. Goodman, \emph{Introduction to Fourier Optics}, 3rd~ed.\hskip 1em plus
  0.5em minus 0.4em\relax Roberts and Company Publishers, 2004.

\bibitem{Salem_OE_06_2014}
M.~A. Salem and C.~Caloz, ``Manipulating light at distance by a metasurface
  using momentum transformation,'' \emph{\jstyle{Opt. Express}}, vol.~22,
  no.~12, pp. 14\,530--14\,543, Jun. 2014.

\bibitem{Achouri_TMTT_07_2015}
K.~Achouri, M.~A. Salem, and C.~Caloz, ``General metasurface synthesis based on
  susceptibility tensors,'' \emph{\jstyle{IEEE Trans. Antennas Propag.}},
  vol.~63, no.~7, pp. 2977--2991, Jul. 2015.

\bibitem{Salem_PIER_11_2014}
M.~A. Salem, K.~Achouri, and C.~Caloz, ``Metasurface synthesis for
  time-harmonic waves: exact spectral and spatial methods,''
  \emph{\jstyle{Progress Electromag. Research}}, vol. 149, pp. 205--216, Nov.
  2014, \istyle{invited}.

\bibitem{Achouri_AXV1510_10_2015_2}
K.~Achouri, G.~Lavigne, M.~A. Salem, and C.~Caloz, ``Metasurface spatial
  processor for electromagnetic remote control,''
  \emph{\jstyle{arXiv}:1510.05726}, Oct. 2015.

\bibitem{Achouri_AXV1510_1_10_2015}
K.~Achouri, B.~A. Khan, S.~Gupta, G.~Lavigne, M.~A. Salem, and C.~Caloz,
  ``Synthesis of electromagnetic metasurfaces: principles and illustrations,''
  \emph{\jstyle{arXiv}:1510.05997}, Oct. 2015.

\bibitem{Kodera_APL_07_2011}
T.~Kodera, D.~L. Sounas, and C.~Caloz, ``Artificial {F}araday rotation using a
  ring metamaterial structure without static magnetic field,''
  \emph{\jstyle{Appl. Phys. Lett.}}, vol.~99, no.~3, pp. 031\,114:1--3, Jul.
  2011.

\bibitem{Sounas_TAP_01_2013}
D.~L. Sounas, T.~Kodera, and C.~Caloz, ``Electromagnetic modeling of a
  magnet-less non-reciprocal gyrotropic metasurface,'' \emph{\jstyle{IEEE
  Trans. Antennas Propag.}}, vol.~61, no.~1, pp. 221--231, Jan. 2013.

\bibitem{Kodera_AWPL_01_2012}
T.~Kodera, D.~L. Sounas, and C.~Caloz, ``Non-reciprocal magnet-less {CRLH}
  leaky-wave antenna based on a ring metamaterial structure,''
  \emph{\jstyle{IEEE Antennas Wirel. Propag. Lett.}}, vol.~10, pp. 1551--1554,
  Jan. 2012.

\bibitem{Kodera_AWPL_12_2012}
------, ``Switchable magnet-less non-reciprocal metamaterial ({MNM}) and its
  application to a switchable {F}araday rotation metasurface,''
  \emph{\jstyle{IEEE Antennas Wirel. Propag. Lett.}}, vol.~11, pp. 1454--1457,
  Dec. 2012.

\bibitem{Kodera_TMTT_03_2013}
------, ``Magnetless non-reciprocal metamaterial ({MNM}) technology:
  application to microwave components,'' \emph{\jstyle{IEEE Trans. Microw.
  Theory Tech.}}, vol.~61, no.~3, pp. 1030--1042, Mar. 2013.

\bibitem{Gupta_TMTT_04_2009}
S.~Gupta, S.~Abielmona, and C.~Caloz, ``Microwave analog real-time spectrum
  analyzer ({RTSA}) based on the spatial-spectral decomposition property of
  leaky-wave structures,'' \emph{\jstyle{IEEE Trans. Microw. Theory Tech.}},
  vol.~59, no.~12, pp. 2989--2999, Dec. 2009.

\bibitem{Liu_EL_11_2002}
L.~Liu, C.~Caloz, and T.~Itoh, ``Dominant mode ({DM}) leaky-wave antenna with
  backfire-to-endfire scanning capability,'' \emph{\jstyle{Electron. Lett.}},
  vol.~38, no.~23, pp. 1414--1416, Nov. 2002.

\bibitem{Otto_TAP_10_2011}
S.~Otto, A.~Rennings, K.~Solbach, and C.~Caloz, ``Transmission line modeling
  and asymptotic formulas for periodic leaky-wave antennas scanning through
  broadside,'' \emph{\jstyle{IEEE Trans. Antennas Propag.}}, vol.~59, no.~10,
  pp. 3695--3709, Oct. 2011.

\bibitem{Otto_AWPL_06_2012}
S.~Otto, A.~Al-Bassam, A.~Rennings, K.~Solbach, and C.~Caloz, ``Radiation
  efficiency of longitudinally symmetric and asymmetric periodic leaky-wave
  antennas,'' \emph{\jstyle{IEEE Antennas Wirel. Propag. Lett.}}, vol.~11, pp.
  612--615, Jun. 2012.

\bibitem{Otto_TAP_04_2014}
S.~Otto, Z.~Chen, A.~Al-Bassam, A.~Rennings, K.~Solbach, and C.~Caloz,
  ``Circular polarization of periodic leaky-wave antennas with axial asymmetry:
  theoretical proof and experimental demonstration,'' \emph{\jstyle{IEEE Trans.
  Antennas Propag.}}, vol.~62, no.~4, pp. 1817--1829, Apr. 2014.

\bibitem{Otto_TAP_10_2014}
S.~Otto, A.~Al-Bassam, A.~Rennings, K.~Solbach, and C.~Caloz, ``Transversal
  asymmetry in periodic leaky-wave antennas for {B}loch impedance and radiation
  efficiency equalization through broadside,'' \emph{\jstyle{IEEE Trans.
  Antennas Propag.}}, vol.~62, no.~10, pp. 5037--5054, Oct. 2014.

\bibitem{Casares_TAP_08_2006}
F.~P. Casares-Miranda, C.~Camacho-{Pe\~{n}alosa}, and C.~Caloz, ``High-gain
  active composite right/left-handed leaky-wave antenna,'' \emph{\jstyle{IEEE
  Trans. Antennas Propag.}}, vol.~54, no.~8, pp. 2292--2300, Aug. 2006.

\bibitem{Kodera_TMTT_04_2009}
T.~Kodera and C.~Caloz, ``Uniform ferrite-loaded open waveguide structure with
  {CRLH} response and its application to a novel backfire-to-endfire leaky-wave
  antenna,'' \emph{\jstyle{IEEE Trans. Microw. Theory Tech.}}, vol.~57, no.~4,
  pp. 784--795, Apr. 2009.

\bibitem{Nguyen_TMTT_07_2010}
H.~V. Nguyen, A.~Parsa, and C.~Caloz, ``Power-recycling feedback system for
  maximization of leaky-wave antennas radiation efficiency,''
  \emph{\jstyle{IEEE Trans. Microw. Theory Tech.}}, vol.~58, no.~7, pp.
  1641--1650, Jul. 2010.

\bibitem{Kodera_TAP_08_2010}
T.~Kodera and C.~Caloz, ``Integrated leaky-wave antenna duplexer{/}diplexer
  using {CRLH} uniform ferrite-loaded open waveguide,'' \emph{\jstyle{IEEE
  Trans. Antennas Propag.}}, vol.~58, no.~8, pp. 2508--2514, Aug. 2010.

\bibitem{Kodera_TAP_10_2010}
------, ``Low-profile leaky-wave electric monopole loop antenna using the
  $\beta=0$ regime of a ferrite-loaded open waveguide,'' \emph{\jstyle{IEEE
  Trans. Antennas Propag.}}, vol.~58, no.~10, pp. 3165--3174, Oct. 2010.

\bibitem{Abielmona_TAP_04_2011}
S.~Abielmona, H.~V. Nguyen, and C.~Caloz, ``Analog direction of arrival
  estimation using an electronically-scanned {CRLH} leaky-wave antenna,''
  \emph{\jstyle{IEEE Trans. Antennas Propag.}}, vol.~59, no.~4, pp. 1408--1412,
  Apr. 2011.

\bibitem{Gutpa_APS_07_2015_2}
S.~Gupta and C.~Caloz, ``Real–time {2-D} spectral-decomposition using a
  leaky-wave antenna array with dispersive feeding network,'' in
  \emph{\cstyle{IEEE AP-S Int. Antennas Propag. (APS)}}, Vancouver, BC, Jul.
  2015, pp. 29--30.

\bibitem{Gupta_arXiv7791_12_2014}
------, ``Spatio-temporal metasurface for real-time 2-{D} spectrum analysis,''
  \emph{\jstyle{arXiv}:1412.7791}, Dec. 2014.

\bibitem{Collin_1992}
R.~E. Collin, \emph{Foundations for Microwave Engineering}, 2nd~ed.\hskip 1em
  plus 0.5em minus 0.4em\relax McGraw-Hill College, 1992, chap.~11.

\bibitem{Salem_AXV02012_05_2015}
M.~A. Salem and C.~Caloz, ``Space-time cross-mapping and application to wave
  scattering,'' \emph{\jstyle{arXiv}:1504.02012}, May 2015.

\bibitem{Taravati_APS_07_2015}
S.~Taravati and C.~Caloz, ``Space-time modulated nonreciprocal mixing,
  amplifying and scanning leaky-wave antenna system,'' in \emph{\cstyle{IEEE
  AP-S Int. Antennas Propag. (APS)}}, Vancouver, BC, Jul. 2015, pp. 639--640.

\bibitem{Dyson_NRB_2015}
F.~Dyson, \emph{Dreams of Earth and Sky}.\hskip 1em plus 0.5em minus
  0.4em\relax New York Review Books, 2015.

\end{thebibliography}
\bibliographystyle{IEEEtran}

\end{document}